\documentstyle[12pt]{article}
\begin{document}
\begin{center}
{\large WAVEPACKETS AND DUALITY IN NONCOMMUTATIVE PLANAR QUANTUM MECHANICS  }\\
\vskip 2cm
Subir Ghosh\\
\vskip 1cm
Physics and Applied Mathematics Unit,\\
Indian Statistical Institute,\\
203 B. T. Road, Calcutta 700108, \\
India.
\end{center}
\vskip 3cm
{\bf Abstract:}\\
Effects of noncommutativity are investigated in planar quantum mechanics in the
coordinate representation. Generally these issues are addressed by converting to
the momentum space. In the first part of the work we show  noncommutative effects
in a Gaussian wavepacket through the broadening of its width. We also rederive results
on $*$-product of Gaussian wavepackets. In the second part, we construct a "Master"
model for a noncommutative harmonic oscillator by embedding it in an extended space.
Different gauge choices leading to different forms of noncommutativity, (such as between
coordinates only, between momenta only or noncommutativity of a more general kind),
can be studied in a unified and systematic manner. Thus the dual nature of these theories
are also revealed.

 \vskip 2cm
\noindent {\bf{Introduction:}} Non-Commutative (NC) Quantum Field
Theories (QFT) \cite{rev} have created a lot of interest in the
High Energy Physics community because of its direct connection to
certain low energy limits of String theory \cite {sw}.
Noncommutativity is induced in the open string boundaries that are
attached to D-branes, in the presence of a two-form background
field. This phenomenon in turn renders the D-branes into NC
manifolds. Stringy effects are manifested in NCQFT framework. The
major advantage of working in the latter is that the basic
structure of QFT in conventional (commutative) spacetime remains
intact and the fundamental (two-point) correlation functions of
QFT are not modified by NC effects. This vital fact emerges from
the construction of the NC generalization of a conventional QFT
where
the products of the field operators in the QFT action are replaced by $*$%
-product (or Moyal-Weyl product) defined below,
\begin{equation}
\hat f(x)*\hat g(x)=e^{\frac{i}{2}\theta_{\mu\nu}\partial_{\sigma_{\mu}}%
\partial_{\xi_{\nu}}}\hat f(x+\sigma )\hat g(x+\xi )\mid_{\sigma =\xi =0}
=\hat f(x)\hat g(x)+\frac{i}{2}\theta^{\rho\sigma}\partial_{\rho}\hat
f(x)\partial_{\sigma}\hat g(x)+~O(\theta^{2}),  \label{1}
\end{equation}
where $\theta_{\mu\nu}$ is conventionally taken as a constant anti-symmetric
tensor. $\hat f(x)$ stands for the NC extension of $f(x)$. {\footnote{$f(x)$ and $\hat f(x)$
 are related in a non-trivial way by the Seiberg-Witten Map when quantum
gauge theories are being considered. However, this is of no concern in the
present work.}}

Exploiting the above rule (\ref{1}), one derives the spacetime
noncommutativity in $x_\mu $-space,
\begin{equation}
[x_{\mu}, x_{\nu}]_{*}= x_\mu * x_\nu - x_\nu *
x_\mu =i\theta_{\mu\nu}.  \label{2}
\end{equation}
In the present work we will only be concerned with spatial noncommutativity in
a plane and so (\ref{2}) reduces to
\begin{equation}
\{x^i,x^j\}=\theta\epsilon^{ij}~;~~\epsilon^{12}=1.  \label{2a}
\end{equation}
Note that in our classical setup we will interpret the noncommutativity as
a modification in the symplectic structure.

From the structure of the $*$-product it is clear that the $\theta$%
-contributions in the quadratic terms in the NC action will be total
derivatives and hence are dropped from the action. For this reason the free
QFTs are not changed under NC extension \cite{rev}.

Effects of noncommutativity are often analyzed in terms of
momentum variables \cite{rev}. This is because the momentum
degrees of freedom (in general) behave in a canonical way, obeying
the symplectic structure,
\begin{equation}
\{p_{i},p_{j}\}=0~;~~\{x_{i},p_{j}\} =\delta_{ij}.  \label{40}
\end{equation}
Here the coordinates are noncommutative given by (\ref{2a}).

In case of NC quantum field theories, NC effects modify only the interaction
vertices when the fields are expressed by their Fourier transforms through
momentum variables. On the other hand, in NC quantum mechanical models, one
replaces \cite{nair,mitra} the original coordinates by a canonical set of
variables carrying a representation of the NC algebra. In terms of this new
set of variables, the NC effects are manifestly present in the Hamiltonian.

However, it is also imperative to know how the NC
effects are reproduced if one pursues with the NC coordinates.
This brings us to the perspective of our work. In the first part
we concentrate on the non-locality induced by noncommutativity. It
has been shown by Gopakumar, Minwalla and Strominger \cite{gms} (see also \cite{szabo}) that if $f$ and
$g$ are wavepackets of small width $\Delta <<\sqrt{\theta}$ then
their $*$-product has a larger region of support
$\frac{\theta}{\Delta}$. In fact this unusual behavior is
responsible for  existence of the NC scalar solitons \cite{gms} in
higher dimensions. The above relation has been derived in the
momentum representation.

We recover the noncommutativity induced fuzzyness through the
broadening of a Gaussian wavepacket {\it{working throughout with
NC coordinates}}. The calculations are done up to the lowest
nontrivial order in $\theta $ ( {\it{i.e.}} $O(\theta ^2 )$) but
it can be generalized in a straightforward way to higher powers of
$\theta $. This is a new result. We also recover the well known
result \cite{gms} that products of Gaussians become defocussed.

In the second part of our paper, we develop a coordinate representation for
NC quantum mechanics, using the NC planar harmonic oscillator as an example.
The problem is that from the ordinary oscillator Hamiltonian,
\begin{equation}
H=\frac{1}{2m}p_{i}p_{i}+\frac{1}{2}x_{i}x_{i},~~i=1,2  \label{3a}
\end{equation}
replacing the products by $\ast $-products one generates the same $H$ as in (%
\ref{3a}) because
$$
x_{i}\ast x_{i}=x_{i}x_{i}+\frac{i}{2}\theta \epsilon _{ij}\delta
_{ij}\equiv~x^2;~~p_{i}\ast p_{i}=p_{i}p_{i}\equiv
p^{2}.
$$
NC effects in coordinate representation can appear in the expectation values where the
action of an operator ${\cal{O}}$ in the position ($x$) basis  is formally expressed
as $\sim<x\mid *{\cal{O}}\mid\psi>$ {\footnote{I thank Dr. B.Muthukumar for pointing
this out.}}. In the  conventional framework \cite{nair,mitra}, this is simulated by
working with a
new set of canonical variables $\{Q_{i},P_{j}\}=\delta_{ij}$ with the identification,
\begin{equation}
x_{i}=Q_{i}-\frac{\theta}{2} \epsilon _{ij}P_{j}~;~~p_{i}=P_{i},  \label{3b}
\end{equation}
so that the NC algebra of (\ref{2a}-\ref{40}) is maintained. The $H$ written in terms of
$(Q,P)$ variables now exhibit the NC  effect through the $\theta $-dependent terms.
However,{
\it{ the new degrees of freedom ($Q_{i},P_{j}$) are obviously
different from the spatial coordinate and momenta}} and this makes
the physical picture unclear.

We, on the other hand, work with the original (coordinate and momentum)
phase space and embed the system in an extended space, in the Batalin-Tyutin
(Hamiltonian) framework \cite{bt}. (For application of the Batalin-Tyutin prescription,
see for example \cite{sg}.) The NC effects manifest themselves
through the embedding procedure. In fact this generalization acts as a
''Master'' model from which different Hamiltonians with  distinct
symplectic structures can be obtained. This clearly demonstrates the duality between
different types of noncommutativity, (such as between coordinates,
between momenta and between both coordinates as well as momenta), that are
induced by different gauge choices in the same Master model. Indeed, one can
choose a specific form of noncommutativity that is suitable for a particular
problem. Similar conclusions have been reached in \cite{rb} in the
Lagrangian framework and we will borrow heavily from this work.

\vskip .5cm
\noindent
{\bf{NC effects in Gaussian wavepackets:}} We will study the  effect of noncommutativity on a
Gaussian wavepacket in the plane given by $exp(-x^2/\Delta ^2)$ where $x^2\equiv x_ix_i$. We
define the NC generalization of an exponential function in the following
way,
\begin{equation}
(e^A)_{NC}=\sum _{n=0}^{\infty}\frac{(A)^n_*}{n!}=1+A+\frac{1}{2}A*A+ ...,
\label{3}
\end{equation}
so that products in the ordinary space are replaced by $*$-products.

As an aside, we note that this definition keeps the plane waves unchanged if
noncommutativity is switched on since,
\begin{equation}
(e^{ik_{j}x_{j}})_{NC}=e^{ik_{j}x_{j}}  \label{4}
\end{equation}
where we have used the identity,
\begin{equation}
(k.x)* (k.x)=(k.x)^{2}+\frac{i}{2}\theta \epsilon
_{ij}k_{i}k_{j}=(k.x)^{2}.  \label{5}
\end{equation}
This might be considered to be in agreement with the fact that the
\textit{free} field theories remain unchanged in their NC
extension and plane waves represents solutions of the free theory
dynamics.

 Alternatively, one might consider the Baker-Campbel-Hausdorf formula in the present case
\begin{equation}
 e^{ik_{j}x_{j}}\equiv e^{i(k_1x_1+k_2x_2)}=e^{ik_{2}x_{2}}e^{ik_{1}x_{1}}e^{-\frac{i}{2}\theta k_1k_2},
 \label{eg1}
 \end{equation}
 from which the NC-generalization yields
 \begin{equation}
 (e^{ik_{j}x_{j}})_{NC}=e^{ik_{2}x_{2}}*e^{ik_{1}x_{1}}e^{-\frac{i}{2}\theta k_1k_2}.
 \label{eg2}
 \end{equation}
 Phase factor induced by the $*$-product will cancel the Baker-Campbel-Hausdorf phase thus
 recovering (\ref{4}) {\footnote{I thank Dr. B.Muthukumar for comments.}}.

Let us start by rederiving the famous result \cite{gms} that NC effects tend to broaden
the product of Gaussian wave packets. Since these wave packets  appear as examples \cite{gms} of the
simplest scalar solitons in NC spacetime, we briefly introduce the solitons as in \cite{gms}
for completeness.

The energy functional of a scalar theory in 2+1-dimensions is \cite{gms}
\begin{equation}
E=\int~d^2x(\frac{1}{2}\partial^{\mu}\phi\partial_{\mu}\phi +\theta V(\phi )),
\label{s1}
\end{equation}
where the coordinates are scaled to dimensionless variables $x\rightarrow x\sqrt{\theta}$.
The polynomial potential in NC spacetime is
$V(\phi )=c(\phi *\phi *...*\phi )$. In the limit $\theta\rightarrow \infty$ only
the potential term in (\ref{s1}) survives and its extremization gives the equation of motion,
\begin{equation}
\frac{\partial V}{\partial \phi}=0.
\label{s2}
\end{equation}
It was pointed out in \cite{gms} that non-trivial solutions $\phi_{0}$ of (\ref{s2}) can
appear provided the following generic relation holds:
\begin{equation}
(\phi_{0}*\phi_{0})(x)=\phi_{0}(x).
\label{s3}
\end{equation}
That the above is possible in NC spacetime was demonstrated in \cite{gms} by considering
a Gaussian wave packet
\begin{equation}
\psi_{\Delta}=\frac{1}{\pi\Delta ^2}e^{-\frac{x^2}{\Delta^2}},
\label{s4}
\end{equation}
and its Fourier transform
\begin{equation}
\tilde \psi_{\Delta}(k)=\int~d^2x e^{ik.x}\psi_{\Delta}(x)=e^{-\frac{k^2\Delta ^2}{4}}.
\label{s5}
\end{equation}
Taking in to account the effect of $*$-product one finds
$$
(\tilde \psi_{\Delta}*\tilde \psi_{\Delta})(p)=\frac{1}{(2\pi)^2}\int~d^2k \tilde \psi_{\Delta}(k)\tilde \psi_{\Delta}(p-k)e^{\frac{i}{2}\epsilon_{\mu\nu}k^{\mu}(p-k)^{\nu}}$$
\begin{equation}
=\frac{1}{2\pi\Delta^2}e^{-\frac{p^2}{8}(\Delta^2 +\frac{1}{\Delta ^2})},
\label{s6}
\end{equation}
which in coordinate space gives
\begin{equation}
(\psi_{\Delta}* \psi_{\Delta})(x)=\frac{1}{\pi^2\Delta ^2(\Delta^2 +\frac{1}{\Delta ^2})}e^{\frac{-2x^2}{(\Delta^2 +\frac{1}{\Delta ^2})}}.
\label{s7}
\end{equation}
Clearly, a packet with $\Delta ^2=1$ squares to itself (modulo a
factor of $2\pi$) \cite{gms}. Note that $\phi_{0}(x)$ is a
solution of the field equation (\ref{s2}) in the
$\theta\rightarrow \infty$ limit but the derivation in
(\ref{s4})-(\ref{s7}) is independent of this restriction.

Now similar results can be derived in our formalism as we now demonstrate. The basic
identity to be exploited now is
\begin{equation}
x^2* x^2=x^{2}-\theta ^{2},
\label{6}
\end{equation}
where we have used
$$
\theta _{ij}\theta _{ij}=\theta ^{2}\epsilon _{ij}\epsilon _{ij}=2\theta
^{2}
$$
considering two spatial dimensions. Utilizing our definition of the $*$-exponential we now
 obtain to $O(\theta ^{2}),$
\begin{equation}
e^{-\frac{x^{2}}{\Delta ^{2}}}* e^{-\frac{x^{2}}{\Delta
^{2}}}\approx e^{- \frac{2\theta ^{2}}{\Delta
^{4}}}e^{-\frac{x^{2}}{\frac{\Delta ^{2}}{2}(1+\frac{ 2\theta
^{2}}{\Delta ^{4}})}}, \label{9}
\end{equation}
which can be compared to the $O(\theta ^2)$ term in (\ref{s7})
\cite{gms}. The outline of derivation is given in Appendix A. Here
NC effect can mask the normal decrease of the resultant width of
the product of Gaussian packets and it is possible to reproduce
the same width in the product Gaussian. This is the reason for the
existence of NC scalar solitons \cite{gms}.

This brings us to our contribution in the present context which is a new result.
It seems natural that if noncommutativity is turned on, a single (Gaussian) wave packet
also will not escape the broadening effect. Our scheme of computation can be applied to a
Gaussian packet
 of width $
\Delta $,
\begin{equation}
f(x)=e^{-\frac{x^2}{\Delta ^2}}.
\label{r}
\end{equation}
It is now straightforward to
derive the $O(\theta ^2)$ modification of the wavepacket using
(\ref{3},\ref{r},\ref{6}). The NC wavepacket to $ O(\theta ^{2})$ is
\begin{equation}
(e^{-\frac{x^2}{\Delta ^{2}}})_{NC}\approx e^{-\frac{\theta ^{2}}{2\Delta ^{4}}
}e^{-\frac{x^2}{\Delta _{NC}^{2}}},  \label{7}
\end{equation}
where the new width is
\begin{equation}
\Delta _{NC}^{2}=\Delta ^{2}(1+\frac{\theta ^{2}}{3\Delta ^{4}}).  \label{8}
\end{equation}
Clearly the new width is larger than the original one indicating
that the sharpness of the packet is reduced. Interestingly, the
correction depends on $ \Delta $ such that the sharper the
original wavepacket, the more fuzzy it becomes. This is one of our
main results which has not been reported before. A few relevant
steps of the computation are provided in the Appendix B.

In the Figure, we have plotted $\Delta _{NC}~vs.~\Delta $ for different values of $\theta$.
The graphs show that $\Delta _{NC}$ saturates for large enough values of $\Delta $ and the
critical value of $\Delta $ increases as $\theta$ increases. For instance
for $\theta ^2 =0.001$ and  $\theta ^2 =1.0$ the saturation values of $\Delta$ are $0.27$
and $1.6$.
This completes the first part of our work.
\vskip .5cm

\noindent
{\bf{NC Harmonic Oscillator:}} In this part of the work our aim is to generalize
the mechanics in NC spacetime in the coordinate representation. As we have
mentioned in the Introduction, it appears that it is not possible to
incorporate the NC corrections in the coordinate representation in a naive
way by elevating the ordinary products to $*$-products. On the other hand,
we wish to avoid the conventional usage \cite{nair,mitra} of a new canonical set of
variables which are different
from spatial coordinate and momentum. We now proceed to show that it is
indeed possible to include NC corrections in the coordinate framework if the
system is embedded in an extended phase space \cite{bt}. We will work with a specific
model for an NC Harmonic Oscillator
\cite{rb},
\begin{equation}
L=q_{i}\dot{x}_{i}+\frac{\theta }{2}\epsilon _{ij}q_{i}\dot{q}_{j}-\frac{k}{2%
}x^{2}-\frac{1}{2m}q^{2}.  \label{10}
\end{equation}
The connecction between $\theta$ and the conventional models of noncommutativity arising
in the coordinates of charged particle moving in a plane with a normal magnetic field is
discussed in \cite{rb}. This first order Lagrangian  possesses the following set of constraints,
\begin{equation}
\psi _{1}^{i}\equiv \pi _{x}^{i}-q^{i}~;~~\psi _{2}^{i}\equiv \pi _{q}^{i}+%
\frac{\theta }{2}\epsilon ^{ij}q^{j}  \label{11}
\end{equation}
and the non-singular nature of the commutator matrix for the constraints
\begin{equation}
\Psi _{\alpha \beta }^{ij}=\{\psi _{\alpha }^{i},\psi _{\beta }^{j}\}~;~~\alpha ,\beta \equiv 1,2
\label{11a}
\end{equation}
where
$$
\Psi _{\alpha \beta }^{ij}=\left(
\begin{array}{cc}
0 & -\delta ^{ij} \\
\delta ^{ij} & \theta \epsilon ^{ij}
\end{array}
\right)
$$
indicates that the constraints are Second Class constraints (SCC) in the
Dirac terminology \cite{dirac} {\footnote{The commuting or First Class Constraints
\cite{dirac} that signal the presence of local gauge invariance will become relevant when
we embed in the extended space \cite{bt}.}}. The
SCCs require a change in the symplectic structure in the form of Dirac
Brackets \cite{dirac}
\begin{equation}
\{A,B\}_{DB}=\{A,B\}-\{A,\psi _{\alpha }^{i}\}\Psi _{\alpha \beta
}^{(-1)ij}\{\psi _{\beta }^{j},B\}.  \label{11b}
\end{equation}
The Dirac brackets are compatible with the SCCs. In the present case, the
inverse of the constraint matrix
$$
\Psi _{\alpha \beta }^{(-1)ij}=\left(
\begin{array}{cc}
\theta \epsilon ^{ij} & \delta ^{ij} \\
-\delta ^{ij} & 0
\end{array}
\right)
$$
leads to the following set of Dirac brackets,
$$
\{x^{i},x^{j}\}_{DB}=\theta \epsilon ^{ij}~,~~\{x^{i},q^{j}\}_{DB}=\delta
^{ij}~;~~\{q^{i},q^{j}\}_{DB}=0,  $$
\begin{equation}
\{q^{i},\pi_q^{j}\}_{DB}=0~;~~\{x^{i},\pi_x^{j}\}_{DB}=\delta
^{ij}~;~~\{\pi_x^{i},\pi_q^{j}\}_{DB}=0.
\label{11c}
\end{equation}
The Hamiltonian is
\begin{equation}
H=\frac{k}{2}x^{2}+\frac{1}{2m}q^{2}.  \label{12}
\end{equation}
Note the spatial noncommutativity and also the fact that $q^i$
behaves effectively as the conjugate momentum to $x^i$. However, the noncommutativity
does not show up explicitly in the Hamiltonian.

We now
wish to reexpress this model in the background  of normal ({\it{i.e.}} commuting) spatial
coordinates which is necessary if one is interested in quantizing the model either in the
Schrodinger or operator formalism. This requires an embedding of the model in a Batalin-Tyutin
extended space
\cite{bt}. (For details of the procedure and
applications see for example \cite{bt,sg}. Here we
only provide the results.)  It is important to remember that the extended space
is {\it{completely}} canonical and the Dirac
brackets of (\ref{11c}) are not to be used. This is
because the SCCs (\ref{11}) are absent in the extended
space and their place is taken by the FCCs that we derive below. \
Introducing a canonical set of auxiliary variables
\begin{equation}
\{\phi _{\alpha }^{i},\phi _{\beta }^{j}\}=\epsilon _{\alpha \beta }\delta
^{ij}~,\alpha ,\beta =1,2~;~~\epsilon _{12}=1,  \label{13}
\end{equation}
it is possible to convert the SCCs in (\ref{11}) to
the following FCCs $\tilde{\psi}_{\alpha }^{i}$,
$$
\tilde{\psi}_{1}^{i}\equiv \psi _{1}^{i}+\phi _{2}^{i}=\pi
_{x}^{i}-q^{i}+\phi _{2}^{i},
$$
\begin{equation}
\tilde{\psi}_{2}^{i}\equiv \psi _{2}^{i}-\phi _{1}^{i}-\frac{\theta }{2}%
\epsilon ^{ij}\phi _{2}^{j}
=\pi _{q}^{i}+\frac{\theta }{2}\epsilon ^{ij}q^{j}-\phi _{1}^{i}-\frac{%
\theta }{2}\epsilon ^{ij}\phi _{2}^{j} , \label{14}
\end{equation}
so the $\tilde \psi ^i_\alpha $ are commutating,
\begin{equation}
\{\tilde{\psi}_{\alpha }^{i},\tilde{\psi}_{\beta }^{j}\}=0.  \label{15}
\end{equation}
Thus the embedded model in extended space possesses local gauge invariance.
For any function $Q$ of the variables, one can construct \cite{bt} its FC extension $\tilde Q$
so
that
\begin{equation}
\{\tilde{Q},\tilde{\psi}_{\alpha }^{i}\}=0.  \label{16}
\end{equation}
This implies that $\tilde O$ is gauge invariant in the extended space. The connection
between the operator algebra in any reduced ({\it{i.e.}} gauge fixed) offspring with its
gauge invariant parent model is the following identity:
\begin{equation}
\{A,B\}_{DB}=\{\tilde A, \tilde B\}.
\label{id}
\end{equation}
In the present case we compute FC counterparts of $x^i$ and $q^i$:
$$
\tilde{q}^{i}=q^{i}-\phi _{2}^{i}~,~~\tilde{x}^{i}=x^{i}-\phi _{1}^{i}+\frac{
\theta }{2}\epsilon ^{ij}\phi _{j}^{2},$$
\begin{equation}
\tilde \pi _x^i =\pi _x^i ~,~~\tilde \pi _q^i =\pi _q^i-\phi^{i}_{1},
  \label{17}
\end{equation}
which in turn generates the FC Hamiltonian
\begin{equation}
\tilde{H}=\frac{k}{2}\tilde{x}^{2}+\frac{1}{2m}\tilde{q}^{2}
=\frac{k}{2}(x^{i}-\phi _{1}^{i}+\frac{\theta }{2}\epsilon ^{ij}\phi
_{2}^{j})^{2}+\frac{1}{2m}(q^{i}-\phi _{2}^{i})^{2}.  \label{18}
\end{equation}
The FCCs take a simpler form,
\begin{equation}
\tilde{\psi}_{1}^{i}= \tilde\pi
_{x}^{i}-\tilde q^{i}~;~~\tilde{\psi}_{2}^{i}
=\tilde\pi _{q}^{i}+\frac{\theta }{2}\epsilon ^{ij}\tilde q^{j}.
\label{ffc}
\end{equation}
This is the cherished form of the Hamiltonian where the NC correction has
appeared explicitly in a fully canonical space with commuting space
coordinates. However we would like to keep the set of dynamical equations
\cite{rb}
\begin{equation}
\dot{q}^{i}=-kx^{i}~;~~\dot{x}^{i}=\frac{1}{m}q^{i}+\theta k\epsilon
^{ij}x^{j}  \label{19}
\end{equation}
unchanged and this can be achieved by adding suitable terms in the Hamiltonian that are
proportional to the FCCs.  We construct $\tilde
H_{Total}$
\begin{equation}
\tilde{H}_{Total}=\tilde{H}+\lambda _{1}^{i}\psi _{1}^{i}+\lambda
_{2}^{i}\psi _{2}^{i}  \label{20}
\end{equation}
and identify the arbitrary multipliers $\lambda ^i_\alpha $ from (\ref{20}):
\begin{equation}
\dot{q}^{i}=\{q^{i},\tilde{H}_{Total}\}=\lambda _{2}^{i}~;~~\dot{x}%
^{i}=\{x^{i},\tilde{H}_{Total}\}=\lambda _{1}^{i}.  \label{21}
\end{equation}
Hence the final form of the gauge invariant FC Hamiltonian generating the
correct dynamics of \cite{rb} is
$$
\tilde{H}_{Total}=\frac{k}{2}(x^{i}-\phi _{1}^{i}+\frac{\theta }{2}\epsilon
^{ij}\phi ^{j}_{2})^{2}+\frac{1}{2m}(q^{i}-\phi _{2}^{i})^{2}
$$
\begin{equation}
+(\frac{1}{m}q^{i}+\theta k\epsilon ^{ij}x^{j})(\pi _{x}^{i}-q^{i}+\phi
_{2}^{i})-kx^{i}(\pi _{q}^{i}+\frac{\theta }{2}\epsilon ^{ij}q^{j}-\phi
_{1}^{i}-\frac{\theta }{2}\epsilon ^{ij}\phi _{2}^{j}).  \label{22}
\end{equation}
This is the ''Master'' model that we advertised in the Introduction
and constitute the other main result of the present paper. The local gauge
invariance in the enlarged space allows us to choose gauge conditions
(according to our convenience) which in turn gives rise to different forms of
symplectic structures and Hamiltonians, that , however, are gauge
{\it{equivalent}}. This is the duality between different structures of noncommutativity,
referred to earlier. As an obvious gauge choice, the unitary gauge
\begin{equation}
\psi _{3}^{i}\equiv \phi _{1}^{i}~,~~\psi _{4}^{i}\equiv \phi
_{2}^{i}  \label{23}
\end{equation}
restricts the system to the original physical subspace with the spatial
noncommutativity as given in (\ref{11c}) being induced
by the full set of SCCs (the original FCCs $\tilde \psi ^i_1 ,\tilde \psi ^i_2$ and the
gauge fixing constraints $\psi ^i_3,\psi ^i_4$).

On the other hand, the following non-trivial gauge
\begin{equation}
\psi _{3}^{i}\equiv \phi _{2}^{i}~;~~\psi _{4}^{i}\equiv \phi _{1}^{i}-\frac{%
\theta }{2}\epsilon ^{ij}\phi _{2}^{j}+c\epsilon ^{ij}\pi _{x}^{j},
\label{24}
\end{equation}
with $c$ being an arbitrary parameter, generates the constraint matrix
\begin{equation}
\Psi _{\alpha \beta }^{ij}=\{\tilde{\psi}_{\alpha }^{i},\psi _{\beta }^{j}\}
\label{25}
\end{equation}
where,
$$
\Psi _{\alpha \beta }^{ij}=\left(
\begin{array}{cccc}
0 & 0 & 0 & -\delta ^{ij} \\
0 & 0 & -\delta ^{ij} & 0 \\
0 & \delta ^{ij} & 0 & -\delta ^{ij} \\
\delta ^{ij} & 0 & \delta ^{ij} & \theta \epsilon ^{ij}
\end{array}
\right) .
$$
The inverse matrix
$$
\Psi _{\alpha \beta }^{(-1)ij}=\left(
\begin{array}{cccc}
\theta \epsilon ^{ij} & \delta ^{ij} & 0 & \delta ^{ij} \\
-\delta ^{ij} & 0 & \delta ^{ij} & 0 \\
0 & -\delta ^{ij} & 0 & 0 \\
-\delta ^{ij} & 0 & 0 & 0
\end{array}
\right)
$$
induces the Dirac brackets,
\begin{equation}
\{x^{i},x^{j}\}_{DB}=(\theta +2c)\epsilon ^{ij}~;~~\{x^{i},\pi _{x}^{j}\}_{DB}=\delta
^{ij}~;~~\{\pi _{x}^{i},\pi _{x}^{j}\}_{DB}=0.  \label{26}
\end{equation}
The choice $c=-\theta /2$ reduces the
phase space to a canonical one with  the Hamiltonian
\begin{equation}
H=\frac{k}{2}x^{2}+(\frac{1}{2m}+\frac{k\theta ^{2}}{8})\pi _{x}^{2}-\frac{
k\theta }{2}\epsilon ^{ij}x^{i}\pi _{x}^{j}.  \label{27}
\end{equation}
Rest of the non-trivial Dirac brackets are not important in the present case.
This structure is in fact identical to the one studied in \cite{mitra}.

Another interesting gauge is
\begin{equation}
\psi _{3}^{i}\equiv q^{i}-\phi _{2}^{i}+ax^{i}~;~~\psi _{4}^{i}\equiv
x^{i}-\phi _{1}^{i}+\frac{\theta }{2}\epsilon ^{ij}\phi _{2}^{j}+bq^{i},
\label{27a}
\end{equation}
where $a$ and $b$ are arbitrary parameters.
Once again, the subsequent constraint matrix
$$
\Psi _{\alpha \beta }^{ij}=\left(
\begin{array}{cccc}
0 & 0 & -a\delta ^{ij} & 0 \\
0 & 0 & 0 & -b\delta ^{ij} \\
a\delta ^{ij} & 0 & 0 & -\delta ^{ij} \\
0 & b\delta ^{ij} & \delta ^{ij} & \theta \epsilon ^{ij}
\end{array}
\right)
$$
and its inverse
$$
\Psi _{\alpha \beta }^{(-1)ij}=\left(
\begin{array}{cccc}
0 & -\frac{\delta ^{ij}}{ab} & \frac{\delta ^{ij}}{a} & 0 \\
\frac{\delta ^{ij}}{ab} & \frac{\theta \epsilon ^{ij}}{b^{2}} & 0 & \frac{%
\delta ^{ij}}{b} \\
-\frac{\delta ^{ij}}{a} & 0 & 0 & 0 \\
0 & -\frac{\delta ^{ij}}{b} & 0 & 0
\end{array}
\right)
$$
results in the symplectic structure,
\begin{equation}
\{x^{i},x^{j}\}_{DB}=0~;~~\{q^{i},q^{j}\}_{DB}=\frac{\theta }{b^{2}}\epsilon
^{ij}~;~~\{x^{i},q^{j}\}_{DB}=-\frac{\delta ^{ij}}{ab}.  \label{29}
\end{equation}
The choice of the parameters $a=\pm 1,b=\mp 1$ fixes $q^i$ to be the
conjugate momentum to $x^i$  but {\it{the momentum variables have
now become noncommutative}}. Comparing the Hamiltonian
\begin{equation}
H=\frac{k}{2}q^{2}+\frac{1}{2m}x^{2}  \label{30}
\end{equation}
with (\ref{12}) (where $x^i$ were noncommutative) one observes that
$q^i$ and $p^i$ have replaced one another. This exercise clearly
demonstrates the fact that coordinate or momentum noncommutativity
are actually different sides of the same coin and are connected by
gauge transformations.

We conclude with a brief comment on the angular momentum $L$ of the system. Remembering that in the extended space, the symplectic structure is completely canonical, $L$ will have the obvious form
\begin{equation}
L=\epsilon^{ij}(x^i\pi^{j}_{x}+q^i\pi^{j}_{q}+\phi^{i}_{1}\phi^{j}_{2}).
\label{l1}
\end{equation}
Upon utilizing the canonical commutators, $L$ will generate
correct transformations on the degrees of freedom. But notice that
$L$ is not gauge invariant ({\it{i.e.}} FC) in the extended space.
Its FC generalization will be,
\begin{equation}
L=\epsilon^{ij}(\tilde x^i\tilde \pi^{j}_{x}+\tilde q^i\tilde \pi^{j}_{q}),
\label{l2}
\end{equation}
since $\tilde\phi^{i}_{\alpha}=0$. To recover the correct transformations of the physical variables one has to construct $L_{Total}$,
$$L_{Total}=L+\xi^{i}_{\alpha}\tilde \psi^{i}_{\alpha},$$
\begin{equation}
\xi^{i}_{1}=\epsilon^{ij}(-\phi^{j}_{1}+\frac{\theta}{2}\epsilon^{jk}\phi^{k}_{2})~;~~\xi^{i}_{2}=-\epsilon^{ij}\phi^{j}_{2}.
\label{l3}
\end{equation}
In the unitary gauge $\tilde L_{Total}$ reduces to the expression given in \cite{rb}. For other gauge choices the structure of angular momentum will be different and the corresponding Dirac brackets will reproduce the transformation.
\vskip 1cm
\noindent
{\bf{Acknowledgements:}} It is a
pleasure to thank Professor J.Gamboa and Professor
A.P.Polychronakos for correspondence and Professor R.Banerjee and
Professor P.Mitra for discussions. Correspondence with Dr. B.Muthukumar is gratefully
acknowledged.
\vskip 1cm
\noindent
{\bf{Appendix A:}} The $*$-product of two Gaussions is
\begin{equation}
e^{-\frac{x^2}{\Delta ^2}}*e^{-\frac{x^2}{\Delta ^2}}=\sum^{\infty}_{n=0,m=0}\frac{(-1/\Delta ^2)^{n+m}}{n!m!}(x^2)^n *(x^2)^m.
\label{b1}
\end{equation}
We have assumed that the NC correction in each Gaussian derived in (\ref{7}-\ref{8}) has
been included in $\Delta $ in (\ref{b1}). Again, to $O(\theta ^2)$, this can be simplified to
$$
e^{-\frac{x^2}{\Delta ^2}}*e^{-\frac{x^2}{\Delta ^2}}\approx \sum^{\infty}_{n=0,m=0}\frac{(-1/\Delta ^2)^{n+m}}{n!m!}[(x^2)^{n+m}-2\theta^{2}nm(m+n-1)(x^2)^{m+n-2}] $$
\begin{equation}
\approx e^{-2\frac{x^2}{\Delta ^2}}-2\frac{\theta ^2}{\Delta ^4}(1-2\frac{x^2}{\Delta ^2})e^{-2\frac{x^2}{\Delta ^2}}\approx
e^{-\frac{2\theta ^2}{\Delta ^4}}
e^{-\frac{x^2}{\frac{\Delta ^2}{2}(1+\frac{2\theta^2}{\Delta ^4})}}.
\label{b2}
\end{equation}
\noindent
{\bf{Appendix B:}} We
consider only up to $O(\theta ^2)$ terms. Using (\ref{6}) it can be
proved that
\begin{equation}
(x^2)^{n}_{*}\approx (x^2)^{n}-\theta^{2}\frac{n(n-1)(2n-1)}{6}(x^2)^{n-2}.
\label{a1}
\end{equation}
The result (\ref{7}-\ref{8}) can be derived using the identity
\begin{equation}
\sum^{n}_{m=1}m^2=\frac{n(n+1)(2n+1)}{6}.
\label{a2}
\end{equation}

\end{document}